\documentclass[preprintnumbers,superscriptaddress,endnote,nofootinbib,preprint]{revtex4}
\usepackage{graphicx}

\usepackage[hypertex]{hyperref}

\newcommand{\lsim}{\lesssim}
\newcommand{\gsim}{\gtrsim}
\newcommand{\lh}{\Lambda_h}
\newcommand{\ee}{e^+e^-}
\newcommand{\snia}{SN$_{\rm Ia}$\,}
\newcommand{\sncc}{SN$_{\rm cc}$\,}

\newcommand{\ord}[1]{\mathcal{O}{(#1)}}
\newcommand{\beq}{\begin{equation}}
\newcommand{\eeq}{\end{equation}}
\newcommand{\eps}{\varepsilon}
\newcommand{\eq}[1]{Eq.~(\ref{#1})}
\newcommand{\Ref}[1]{Ref.~\cite{#1}}

\begin{document}

\pagestyle{plain}


\title{The INTEGRAL/SPI 511 keV Signal from Hidden Valleys\\ in \\
Type Ia and Core Collapse Supernova Explosions}

\author{Hooman Davoudiasl\footnote{email: hooman@bnl.gov}}
\affiliation{Department of Physics, Brookhaven National Laboratory,
Upton, NY 11973, USA}

\author{Gilad Perez\footnote{email: Gilad.Perez@weizmann.ac.il}}
\affiliation{Department of Particle Physics \& Astrophysics, Weizmann Institute of Science, Rehovot 76100, Israel}


\begin{abstract}

We examine under what circumstances the INTEGRAL/SPI 511 keV signal
can originate from decays of MeV-scale composite states produced by:
(A)  thermonuclear (type Ia) or (B) core collapse supernovae (SNe).
The requisite dynamical properties that would account for the
observed data are quite distinct, for cases (A) and (B).  We
determine these requirements in simple hidden valley models, where
the escape fraction problem is naturally addressed, due to the long
lifetime of the new composite states.  A novel feature of scenario
(A) is that the dynamics of type Ia SNe, standard candles for
cosmological measurements, might be affected by our mechanism.  In
case (A), the mass of the state mediating between the hidden sector
and the SM $\ee$ could be a few hundred GeV and within the reach of
a 500 GeV $\ee$ linear collider.  We also note that kinetic mixing
of the photon with a light vector state may provide an interesting
alternate mediation mechanism in this case.  Scenarios based on case
(B) are challenged by the need for a mechanism to transport some of
the produced positrons toward the Galactic bulge, due to the
inferred distribution of core collapse sources.  The mass of the
mediator in case (B) is typically hundreds of TeV, leading to
long-lived particles that could, under certain circumstances,
include a viable dark matter candidate. The appearance of long-lived
particles in typical models leads to cosmological constraints and we
address how a consistent cosmic history may be achieved.

\end{abstract}
\maketitle


\section{Introduction}

The recent INTEGRAL/SPI measurement of the 511 keV signal~\cite{Knodlseder:2005yq} from
$\ee$ annihilation in the Milky Way reconfirms~\cite{Teegarden:2004ct}
its puzzling characteristics: it appears that $\ee$
pairs annihilate at a rate of $1.5\pm 0.1 \times 10^{43}$\,s$^{-1}$ in the Galactic
bulge, whereas the corresponding rate in the disk is $0.3\pm 0.2 \times 10^{43}$\,s$^{-1}$.
Hence, the bulge-to-disk annihilation ratio $B/D \sim 3-9$ is implied~\cite{Knodlseder:2005yq}.
The size of the signal and the inferred value of $B/D$ make it a
challenge to explain the data using standard
astrophysical sources.

A leading conventional candidate would be the
radioactive $\beta^+$ decays of type Ia Supernova
(SN) products, such as $^{56}$Co.
These explosions can supply the necessary flux of $e^+$ and
are thought to happen mostly in the central region of the galaxy,
with an old stellar population~\cite{Knodlseder:2005yq}.
However, it has been estimated that the fraction of the
$e^+$ that can escape the type Ia ejecta is not large enough
to produce the entire observed signal~\cite{Kalemci:2006bz}.

It has also been argued that the bulge rate of SN type Ia (\snia)
explosions can be inferred by mass-scaling of the rate in early type galaxies that
host old stars~\cite{Prantzos:2005pz}; this would then lead to values of
$B/D$ that are too small~\cite{Knodlseder:2005yq} (however, this extrapolation
ignores the differences between the evolutions of the Milky Way bulge and such galaxies).
In this case, unless one assumes that $\ee$ annihilation takes place far (on galactic scales)
from the source~\cite{Prantzos:2005pz}, this ratio cannot be straightforwardly explained.

Various other sources, such as low mass X-ray binaries~\cite{Knodlseder:2005yq},
have been proposed to explain the 511 keV signal.  However, generally speaking,
all such explanations suffer from rather sizable uncertainties and it is not clear if any plausible
astrophysical mechanism can account for the data.  This situation has
motivated new physics proposals, such as those based
on the Galactic dark matter (DM) content; see for example~\cite{Boehm:2003bt,xdm,thdm}.

Here, we study a different set of new physics possibilities that
do not depend on cosmic relic abundances of
particle populations, but rely only on supernova production of new
composite states that decay into $\ee$.  We consider two distinct scenarios,
referred to hereafter as cases (A) and (B):
\begin{itemize}
\item In case (A), the new states are produced in type Ia explosions.
\item In case (B) we
consider production in core collapse (cc) supernovae.
\end{itemize}

A simple estimate
shows why this may, in principle, be energetically possible.
Roughly speaking, at a rate of 1--2 explosions per
century, ${\cal O} \left(10^{48-50}\right)$~MeV s$^{-1}$ of binding energy is
released into the Galaxy.  Obviously, even if a tiny fraction of
this energy flux is carried by the hidden sector, it would be enough
to account for the $\ee$ annihilation signal.
In fact, it was proposed long ago that the 511 keV
signal could have been generated by the decays of heavy
$\tau$-neutrinos (now ruled out) produced in cc SN (\sncc) explosions~\cite{Dar:1986wb}.

As mentioned above, the  511 keV data
could have reasonable standard explanations~\cite{HLR}.
Here, we do not attempt to refute possible
standard astrophysical descriptions of the data.   However, given the uncertainties
involved in these (or other) explanations, we believe alternative proposals merit attention.
In particular, the long lifetime of particles in the scenarios
we consider can naturally address the
escape fraction problem present in the  conventional \snia picture.  Also, our models
could have testable
predictions for collider experiments [case (A)], and may also provide
DM candidates [case(B)], though we do not {\it require} that our scenarios give rise
to a viable DM particle. We note in advance that, while the new scenarios we examine provide
advantages over the conventional accounts, they are subject to their own challenges and constraints.
Nonetheless, they offer concrete and interesting examples of how hidden valley
dynamics could affect astrophysical phenomena, such as the Galactic 511 keV signal.

The 511 keV signal suggests that $\sim 10^{43}$ $\ee$ pairs are
annihilated each second, within the bulge at the center of the
Galaxy.  Astronomical observations require that $e^+$ be injected
into the interstellar medium (ISM) near threshold, at energies below
a few MeV per $e^+$~\cite{Jean:2005af}. We will propose that this
injected flux comes from the decay of a new state $X$ of mass
$m_X\sim 1$~MeV, from a hidden dynamical sector of the type
discussed in Ref.~\cite{hv1}.  We further assume that $X$ only
decays through SM final states and its branching ratio into $\ee$ is
typically $\ord{1}$. Given the complexity of the astrophysical
system and the fact that the hidden dynamics is non-perturbative our
analysis is only done in a semi-quantitative manner. Yet, the
analysis presented below is concrete enough to make the point that
hidden valley models may play an important role in dynamics of
astrophysical systems and the cosmological evolution of the
universe.

In case (A), we find that a successful scenario based on \snia
requires a mediator of mass $\ord{300}$\,GeV,
connecting the SM and the hidden sectors, making its discovery at a 500\,GeV $\ee$ collider a possibility.
We also show how present experimental constraints may be addressed.  Here, we also briefly discuss how
kinetic mixing of the photon with a light vector can provide another potential mediation mechanism between
the hidden and the SM sectors \cite{kinmix}.  The production of the
$X$ particle is Boltzmann suppressed, hence it is not relativistic and its $\ee$ decay
products have MeV-scale energies, as required by observations.  The lifetime of $X$ is long enough to
exit the SN ejecta before decaying and avoids the conventional \snia positron escape fraction problem.  The \snia
Galactic population is centrally peaked and helps explain the spatial distribution of the
511 keV signal. Potentially
troublesome long-lived particles produced in the early universe can be efficiently annihilated
via strong dynamical processes within the hidden sector.
A major concern in this scenario, apart from direct constraints on the model, is whether
Big Bang Nucleosynthesis (BBN) can go through without much perturbation.  Although we show that
this may be plausibly achieved within the considered framework, perhaps a more detailed analysis is
required to reach a firm conclusion.

In case (B), the $X$ production in \sncc is not Boltzmann suppressed and can be brought down to the required levels by
lowering the coupling of the hidden and SM sectors.  This is achieved by assuming a high SM-hidden mediation scale of
$\ord{100}$~TeV, which leads to a long lifetime for $X$.  This feature allows the $\ee$ decay products
to be deposited far from the \sncc sources that are mostly found in the Galactic disk.
This long decay
length could lead to acceptable $B/D$ values, since it causes the surface brightness of the disk $\ee$ annihilations
to be reduced and may also help with magnetic transport of
some positrons into the bulge.
Even though the kinetic energy of the initially produced hidden ``partons" is $\ord{10}$\,MeV, fragmentation
into $X$ and other hidden hadrons softens the final state $\ee$ energy to the desired few MeV level.  Here,
acceptable cosmic abundance of long-lived particles may require low reheat temperatures, but
simple models give rise a to a plausible DM candidate.
The main shortcoming in this scenario stems from the fact that the disk population of the \sncc is
bigger than that of the bulge and a Galactic mechanism is needed to transport the generated positrons into the bulge, as we will
discuss.

In the following section, we discuss the microscopic structure of our models.
In section~\ref{caseA}, we explain how the 511\,keV signal can be addressed
via type Ia supernova explosions. Section~\ref{caseB} deals with the core
collapse case.  In section~\ref{cosmology}, we discuss the cosmological implications of our setup.
Section~\ref{summary} contains our final conclusions.

\section{Microscopic Setup}\label{micro}
We assume that the hidden sector has a dynamical
scale $\lh \sim 1$~MeV at which the new states emerge.  This scale sets the masses
of the new hidden hadrons\footnote{For simplicity,
we do not consider the possibility of having very light pseudo-Goldstone bosons (PGBs) well below $\lh$. Inclusion of
PGB states would further enrich the structure of the theory and would add new scales to the system, hence it
is worth future study.}.
The decays of the composite states then yield the requisite $e^+$ input at $\ord{\rm MeV}$ energies .
Of course, if $X$ is emitted at energies much larger than its mass,
the positrons will be very relativistic, in conflict with observational evidence that
demands a cold positron flux.  However, below we will argue
that Boltzmann suppression, in case (A), or
parton shower and fragmentation, in case (B), makes this unlikely, given the relevant energies.
We will denote the fermionic ``quark" degrees
of freedom in the hidden sector, subject to non-trivial hidden dynamics, by $f$.
Below the scale $\lh$, strong dynamics leads to the emergence of
hidden hadrons made up of $f$ quarks.

Following Ref.~\cite{hv1},
we will assume that the hidden dynamics is coupled to the SM via higher dimension
operators suppressed by a scale $M$.  For concreteness, we will consider the following
dimension-6 operator that can arise in a variety of models:
\beq
\frac{({\bar \psi_i}\Gamma_1 \psi_j)({\bar f_a}\Gamma_2
f_b)}{M^2},
\label{hSM}
\eeq
where $\psi_i$ is an SM fermion and
$\Gamma_{1,2}$ represent a product of Dirac matrices.
For example, if the visible and the hidden
sectors couple through a massive scalar or vector $\Phi$ of mass $m_\Phi$,
then one gets $M\sim m_{\Phi}/\sqrt{g_1 g_2}$~\cite{hv1},
where $g_{1}$ and $g_2$ denote
couplings of the $\Phi$ to the SM and hidden sectors, respectively.  In what follows, we
focus on the case where $\Phi$ is a vector particle.   In our discussion of case (A), we will
briefly consider how the kinetic mixing of a light vector particle with the photon
\cite{kinmix} can provide another mediation mechanism between the hidden and the SM sectors.

In places where a concrete model is appropriate to consider we will adopt the One-Light-Flavor (1LF)
model discussed in \Ref{hv1}.  We will denote the lightest pseudo-scalar by $\eta_h$ and
the lightest vector by $\omega_h$, with $m_{\omega_h}\sim \sqrt{n_c^h} \,m_{\eta_h}$~\cite{hv1},
based on the results from Ref.~\cite{Witten:1978bc}; $n_c^h$ is the number of hidden colors.
There is also a scalar, $\sigma_h$,
that is typically intermediate in mass and
lifetime~\cite{hv1}.  The generic expectation is that
$\sigma_h$ is up to $30\%$ heavier than
$\eta_h$~\cite{hv1}, and hence it would not decay into $\eta_h+\ee$,
within our typical parameter ranges.  Then
the key aspects of the phenomenology we are interested in can be captured by considering
only $\eta_h$ and $\omega_h$, as we do for simplicity in the following.
We also briefly discuss below the possible implications of the
$\sigma_h\to \eta_h \gamma \gamma$ loop-mediated decay.

\section{Case (A): Type Ia Supernova Explosions}\label{caseA}

Let us begin with an estimate of the $X$-production in
\snia.  These events are
attributed to the explosion of accreting white dwarfs through
thermonuclear reactions that disrupt the star entirely, releasing
$\sim 1 \times M_\odot \approx 2\times 10^{33}$~g of binding energy in the process.
Observations of similar processes in other galaxies, including those of novas that
are believed to share the same progenitor, point to a centrally peaked distribution
of \snia~\cite{Knodlseder:2005yq}.  It is then plausible to assume that
they could give rise to $B/D\gsim 1$, as
long as one can provide a way for a large enough fraction of
the generated $e^+$ flux to escape the explosion and reach the interstellar medium (ISM).  We will base our
estimates on the carbon deflagration model of \snia, a detailed discussion of which
can be found in Ref.~\cite{Nomoto:1984sm}; we will adopt the parameters
of the widely used W7 model.

The \snia explosion occurs over about 1\,s, during which a large fraction
of the white dwarf mass, roughly $0.7 M_\odot$, reaches temperatures
of order $T_{\rm Ia} \approx  6\times 10^9$\,K.
We will assume a mean density
$\rho \approx 3\times 10^8$\,g\.cm$^{-3}$~\cite{Nomoto:1984sm}.  For
the electrons in the white dwarf, bremsstrahlung processes in which
an electron scatters from the electromagnetic filed of the
nucleus $N$ are the dominant process:
$e N \to e N X$~\cite{Raffelt}.  The rate for scalar or vector
emission from non-relativistic and non-degenerate plasmas has been
calculated in Ref.~\cite{GMP}.  Since $T_{\rm Ia} \lsim m_e$, and degeneracy
is mild for our reference parameters~\cite{Raffelt}, we adapt their
result for a reasonable estimate of energy loss rate, in ${\rm erg}~{\rm g}^{-1}~ {\rm s}^{-1} $,
\cite{GMP,Raffelt}
\beq
\eps_X \approx e^{-m_X/T_{\rm Ia}} \alpha' \, \eta \,
2.8\times 10^{26} \,
T_8^{0.5}\, Y_e\, \rho \sum_j \frac{X_j Z_j^2}{A_j},
\label{epsX}
\eeq
where $\alpha' \equiv g'^2/(4 \pi)$,
with $g'$ the coupling of the scalar (massive vector) $X$ state
to the electron, corresponding to $\eta = 1 (3)$.  Here, $T_8 \equiv T/(10^8\,{\rm K})$,
$Y_e\simeq 0.5$ is the electron number fraction relative to baryons, $X_j$ is
the mass fraction of species $j$ with atomic number $Z_j$ and mass $A_j$;
we will take $\sum_j X_j Z_j^2/A_j \sim 14$ for our estimates.  In the above formula,
we have introduced a factor for Boltzmann suppression, since $T_{\rm Ia} < m_X$.
We require that the
above rate yield $\sim 10^{43}$ $X$ particles per second, in order to generate
the inferred $\ee$ annihilation signal.  Taking the \snia rate to be roughly 1 per century
in the Galaxy and $m_X = 2.0$\,MeV, we get
\beq
\alpha' \sim 4.6 \times 10^{-22}/\eta\, .
\label{alpha'}
\eeq
With $\eta = 3$ for a vector $X$ and a naive estimate $g' \sim m_X^2/M^2$, we
find $M \sim 300$\,GeV; here and throughout this work we will
assume that the decay constant of a hidden hadron  $f_X\sim m_X$.

A rough estimate of the $X$ lifetime in this scenario is
\beq
\tau_X \sim  \frac{16 \pi}{m_X g'^2} \sim \frac{16 \pi M^4}{m_X^5} \sim 10\,{\rm s}
\left(\frac{M}{300~{\rm GeV}}\right)^4
\left(\frac{2~\rm MeV}{m_X}\right)^5 \, .
\label{tauX1}
\eeq
The size of $\tau_X$ also provides a way to address
the problem of $e^+$ escape fraction.  To see this, we argue that the $X$
particles will typically fly out with a velocity $v_X$, much faster than the
stellar explosion which moves at a relatively low speed $v_{\rm Ia} \sim
0.03$~\cite{Nomoto:1984sm}.
The reason for this is that the energy distribution of the emitted $X$ particles inherits the
electron energy distributions inside the \snia, hence we expect
\begin{equation}
v_X\sim (T_{\rm Ia}/m_X)^{1/2} \gg v_{\rm Ia}\,.
\end{equation}

Finally, we note that the size of the progenitor of a \snia
is given by the radius of the accreting white dwarf, $R_{\rm wd}
\sim 10^4~{\rm km} \sim 0.03$~s.  Then, \eq{tauX1} implies that the
$X$ particles will decay well outside the stellar explosion front
and will not be absorbed by the ejecta.  This avoids the escape
fraction problem of the conventional \snia picture, where the
positrons can get absorbed efficiently as they are released within
the outward moving stellar matter.

\subsection{Direct constraints}
The mass scale $M$ may seem unacceptably low, in light of
the existing LEP data~\cite{LEP, LEPaleph,LEPdelphi},
or direct detection data~\cite{TevatronZ'}.
However, we argue that simple assumptions about the nature
of the coupling of the hidden and SM sectors to the mediating particle $\Phi$ (such as a
heavy $Z'$)  can allow for such values of $M$.  In particular, if we assume that the hidden
sector couples to $\Phi$ strongly with $g_2 \sim 10$, whereas the SM couples to
$\Phi$ weakly with $g_1 \sim 1/g_2$, then many existing
bounds can be satisfied for $m_\Phi \sim 300$~GeV.
As we do not
require $X$ production from SM hadrons in \snia, a simple assumption would be to take
the $Z'$ coupling to quarks to be negligible, in which case the
Tevatron bounds~\cite{TevatronZ'} would not apply.  Here we would like to mention that
kinetic mixing of the photon with a light vector particle can offer
another interesting possibility.  For example, if the mixing is governed by a loop-level
mixing parameter $\varepsilon \sim 10^{-3}$, for $g_2 \sim 0.1$ and
$m_\Phi \sim 1$~GeV, we may achieve the required level of coupling
between the hidden and the SM sectors.  Such parameters can accommodate
precision measurements of $(g-2)_e$ \cite{lowseesaw},
while potentially accounting for a possible deviation
of $(g-2)_\mu$ from the SM prediction \cite{Passera:2008jk}.  However,
for concreteness we will concentrate on the case with $m_\Phi\sim 300$~GeV, in what follows.

The LEP bounds on $\left(\ee\right)_{V+A} \to \left(\ee\right)_{V+A} $, from
Ref.~\cite{LEP}, roughly require
$\Lambda \gsim 7-9$~TeV (depending on the sign of the effective
dimension-6 operator), where $\Lambda = \sqrt{2 \pi} \, m_\Phi/g_1$.  For $g_1\approx 1/10$
and $m_\Phi \approx 300$~GeV we then get $\Lambda \approx 8$~TeV which is in the
realistic regime.  We hence adopt a minimal case where
$\Phi$ couples dominantly to $\left(\ee\right)_{V+A}$ in the SM.  We note that this would introduce
a suppression by a factor of 1/2 in \eq{epsX}, due to the random polarization of the
electrons in the star.  Even then, $M\sim 300$~GeV still yields about $7\times 10^{42}$ $\ee$ pairs
per second which, within the accuracy of our analysis, is of the right size.

Another possible LEP bound comes from $\ee \to \gamma + E\!\!\!\!\slash\,$~\cite{LEPaleph,LEPdelphi}, where
the missing energy is carried by the invisible ${\bar f}f$ jets.
We estimate the cross section for
$\ee \to \gamma {\bar f}f$\, by
\beq
\sigma_{ef\gamma} \approx {1\over 4}\times{\alpha\over \pi}\times \frac{n_c^h E_{\rm cm}^2}{12 \pi M^4}\,,
\label{sigma}
\eeq
where the factor of (1/4) accounts for chiral couplings
of the initial and final states, the factor of $\alpha/\pi$ accounts for the initial state radiation, and
$E_{\rm cm}$ is the center of mass energy.  In deriving the above formula, we have simply
adapted similar results, say, for $\ee \to \mu^+ \mu^-$~\cite{P&S}.  An analogous process arises
in models with large extra dimensions~\cite{ADD}, where the missing energy is carried off by
Kaluza-Klein gravitons~\cite{GRW,MPP}.  Using the results of Ref.~\cite{MPP},
we then infer that $M\approx 300$~GeV is a safe suppression scale, for $E_{\rm cm} \approx
200$~GeV and $n_c^h=2$,
given the LEP2 bounds~\cite{LEPaleph,LEPdelphi} on $\ee \to \gamma + E\!\!\!\!\slash$\ .

\subsection{Astrophysical constraints}
The above inferred value of $\alpha'$ is not constrained by astrophysical processes
at $T \ll m_X$, such as those of red giant cores at $T\sim 10$\,keV.  However,
\sncc explosions, characterized by $T\sim 30$\,MeV, could overproduce
the 511 keV signal and, in principle, lead to severe
constraints on our \snia-based scenario.
Remarkably, for $M\sim 300$\,GeV, we roughly infer a mean free path
$\lambda \sim M^4/T^5 \sim 10^2$\,m for the hidden sector $f$ quarks (even though we are
assuming that the hidden sector has negligible couplings to SM hadrons of the core,
we may expect that such a hot environment provides a photon and $\ee$ thermal bath that can efficiently
generate a population of $f$-quarks).  Since
the size of the hot SN core is $10-100$~km, we see that the $f$ quarks are quite likely trapped
and can only be surface-emitted.  This avoids constraints coming from over-cooling
of the \sncc core, which would otherwise halt the explosion.

One may also worry that
a trapped ``hidden-sphere" will contain a substantial fraction of the explosion
energy and overproduce the 511 keV signal.
However, given the estimation of the $X$ lifetime in Eq.~(\ref{tauX1}) we claim that
over-production is not a worry.
The reason is that the progenitors of the \sncc are red or blue giant stars of radius
$\sim 10^2 R_\odot \sim 10^2$~s, much larger than the decay length of $X$.  Hence,
the $\ee$ decay products of the $X$ particles will be absorbed well-within the \sncc
progenitor and will not contribute to an over-population of $e^+$ in the ISM.  In fact, this effect might
help to prevent the shock wave (which drives the cc explosion) from being stalled,
a major outstanding problem in understanding \sncc.

Regarding $\sigma_h$, we note that it is expected to be at most $30\%$ heavier than
$\eta_h$~\cite{hv1}.  Hence, within our typical parameter range, it would decay into
$\eta_h\,\gamma\gamma$, via a one-loop radiative process,
with an approximate life time
\beq
\tau_\sigma\sim 2\times10^{11}{\rm \,s}\,
\left({0.4 \,{\rm MeV}\over\Delta m_h}\right)^5\,
\left({M\over  300 \,\rm GeV}\right)^4\,,
\eeq
where $\Delta m_h=m_{\sigma_h}-m_{\eta_h}\sim 0.4 \,$MeV.
This is roughly consistent with the bound~\cite{Raffelt}
produced by the Solar Maximum Mission (SMM) satellite
\cite{Chupp:1989kx} looking for a flash
of photons emitted from the 1987a \sncc, yet may require $\Delta m_h$ to be a factor
of few smaller.


\section{Case (B): Core Collapse Supernovae}\label{caseB}

As noted before, \sncc can also produce the requisite
$e^+$ flux to account for the size of the 511 keV signal.  However, as we will see,
this can be done in a very different regime of hidden sector models,
due to the much higher temperature of such explosions.
The estimated rate of energy release from the \sncc, for particles that free stream
out of the core, can be obtained from that of $\nu {\bar \nu}$ emission
from a non-degenerate stellar core, via
$N N \to N N  \nu {\bar \nu}$, with $N$ a nucleon~\cite{Raffelt}
\beq
\eps_\nu = 2.4\times 10^{17} {\rm \,erg}\,{\rm g}^{-1}\,{\rm s}^{-1} \,
\rho_{15} \, T_{\rm MeV}^{5.5}\, .
\label{epsnu}
\eeq
Here, the density $\rho_{15}$ is in units of $10^{15}$\,g/cm$^3$ and
the temperature $T_{\rm MeV}$ is in MeV.  In the \sncc core, $T\sim 30$\,MeV and
the rate saturates at $\rho_{15} \approx 0.03$~\cite{Raffelt}, and we find
$\eps_\nu \approx 9.6\times 10^{24} {\rm \,erg}\,{\rm g}^{-1}\, {\rm s}^{-1}$.
The temperature of the core drops over a time scale of order a few seconds.  Given the
steep temperature dependence of the rate in \eq{epsnu}, we estimate the total $f$-quark
(hidden sector) output from the emission over the first few seconds, by rescaling the
Fermi constant $G_F \simeq 1.2 \times 10^{-5}$\,GeV$^{-2}$ of the SM weak interactions.
For an object of mass $\sim 1.5 M_\odot \sim 3 \times 10^{33}$\,g, the ratio $R_f$
of $f$-quark output to the total energy budget $\sim 3\times 10^{53}$\,erg
is given by
\beq
R_f \sim 10^4(M^2 G_F)^{-2}\,.
\label{Ff}
\eeq
Based on our preceding discussion, $R_f \sim 10^{-6}$ is a reasonable
estimate for the required output.  We then get
$M\sim 100$\,TeV as a natural scale for the operator in \eq{hSM}, if one
wants to explain the 511 keV signal from \sncc production.

Once the $f$-quarks are produced, they would ``hadronize"
on their way out of the SN core.  Showering and fragmentation of the $f$-quarks
will then distribute their emission energy among several $X$ and other hidden
hadrons.  As a rough guide, we will consider
the average multiplicity $N_h$ of
QCD hadrons in $\ee$ collisions.  Given the typical
SN temperature of about 30\,MeV and $\lh \sim 1$\,MeV,
the data at $\sqrt{s}\sim 30$\,GeV will provide a good analogue, and we find
$N_h \approx 20$~\cite{PDG}.  We then expect $\ord{10}$ hidden hadrons
to result from $f$-quark emissions at $T \sim 30$\,MeV.  As long as
the initial emitted parton fragments into at least a few leading hadrons, their energies
will typically be below $\sim 10$\,MeV, which will eventually
yield positrons at few MeV energies, as required by the data.  This feature is
a universal aspect of generic strong dynamics, regardless of the specific model.
Looking at a compilation of measurements for center of mass energies between 14-44\,GeV,
by the Tasso collaboration~\cite{Braunschweig:1990yd}, we find, indeed, that in
$\ee\to q\bar q$ processes less than $1\%$ of the events contain a particle which carries $50\%$
or more of the beam energy and the majority of final state hadrons are soft.

A reasonable fraction of these hidden sector
hadrons must  decay well-outside the \sncc progenitor star, if one wishes to disperse
the $\ee$ throughout the ISM.
Since the typical radii of the collapsing stars is $\ord{10^2 R_\odot}$, we require
$\tau_X \gsim 100$\,s.  Note that the naive expectation for the lifetime of $X$
in this scenario is now
\beq
\tau_X \sim  \frac{16 \pi M^4}{m_X^5} \sim 10^{11}~{\rm s}
\left(\frac{M}{100~{\rm TeV}}\right)^4
\left(\frac{2~\rm MeV}{m_X}\right)^5\,,
\label{tauX2}
\eeq
which is more than sufficient to satisfy this requirement.
As we will discuss below,
this long lifetime could be a useful feature in explaining the distribution of
the 511 keV signal.

The inferred $\ee$ flux seems to come mainly from the Galactic bulge and is not
evenly distributed throughout the Galactic disk.   However, the \sncc events
are expected to be mostly concentrated in the disk.  The Galaxy can be
roughly described as a disk of radius $r_G \approx 15$\,kpc.
We approximate the central bulge of the Galaxy as a
sphere with a radius $r_B \sim 3$\,kpc.  A reasonable lower bound
on the fraction $f_B$ of the cc events that come from the bulge region is obtained by
the ratio of the disk area inside and outside the bulge~\cite{AG-Y}
\beq
f_B \gsim r_B^2/(r_G^2 - r_B^2)\,,
\label{fB}
\eeq
which yields $f_B \gsim 0.04$.  One can also get estimates of the $f_B$ using observational
data on neutron stars and pulsars~\cite{Mirizzi:2006xx}.  The distribution $n_{cc}$ of
\sncc in our galaxy can be described,
in galactocentric cylindrical coordinates $(r, z , \theta)$, by~\cite{Mirizzi:2006xx,Lazauskas:2009yh}
\beq
n_{cc}(r) \propto r^\xi e^{-r/u}\left[0.79\, e^{-(z/0.212)^2} + 0.21\, e^{-(z/0.636)^2}\right],
\label{ncc}
\eeq
where $\xi$ and $u$ are fit parameters, and $z$ is in kpc.
The neutron star data yield $\xi = 4$ and $u = 1.25$~kpc,
whereas the pulsar distribution implies $\xi=2.35$ and
$u=1.53$~kpc~\cite{Mirizzi:2006xx}.  We find that the first set of
parameters yields $f_B \approx 0.04$, whereas
the second set gives $f_B\approx 0.11$.
In either case, we need to address the paucity of bulge sources.

Ref.~\cite{Prantzos:2005pz} has suggested
that if positrons can propagate away from the Galactic disk they will fill out a larger volume and
reduce the surface brightness of the disk annihilation emission.  Also, once the positrons get out
to $z \gsim 3$~kpc, they will escape beyond the cosmic ray halo (CRH), where
Galactic (poloidal) magnetic fields may transport them
into the bulge.  Ref.~\cite{Prantzos:2005pz} argues that, if these conditions are satisfied,
then $B/D$ as low as $\sim 0.5$ can be consistent with the current INTEGRAL/SPI observations.
However, a naive estimate of the escape fraction $f_{\rm esc}$ from the CRH,
using slow-down and confinement time scales for positrons,
yields $f_{\rm esc} \sim 0.1$~\cite{Prantzos:2005pz}.  This is
not sufficient to get the required level of disk $e^+$ population into the bulge,
yet there may be room for enhancing $f_{\rm esc}$, given the uncertainties involved.
We will next discuss how in our scenario $f_{\rm esc} \gg 0.1$ can be naturally
obtained.  This would easily allow us to take advantage of
the mechanism suggested in Ref.~\cite{Prantzos:2005pz}.

The long lifetime $\tau_X$ in \eq{tauX2} is helpful in releasing the requisite
$\ee$ away from the cc explosion and the Galactic disk.
Here, we give a rough estimate for the fraction of the $\ee$
trapped within the CRH.  We only consider $\ee$ from
decays of the $X$ particles in our scenario.  For simplicity, we approximate the CRH
by a slab of thickness $d$ and assume that the planar extent
of the slab (Galactic disk) is large compared to other length scales.  Furthermore,
it is also assumed that the \sncc explosions are confined to a thin layer (set by the thickness
of the Galactic disk) in the middle of the slab,
which is a reasonable approximation.
The fraction of the $X$ particles that decay
within the slab is at most of order
\beq
F_{X}\sim \frac{2}{\pi} \int_0^{\pi/2} \!\!d\theta \,
(1-e^{-l/\tau_X}),
\label{FX}
\eeq
where
$
l = d/(2 \cos\theta).
$
Guided by \eq{tauX2},
let us assume that  $\tau_X \sim 10^{12}$\,s, equivalent to a decay length of
order 10\,kpc; this corresponds to $m_X\approx 2 m_e$.
For $d=6$\,kpc, we then find $F_X \sim 0.5$; for $\tau_X$ equivalent to
20\,kpc ($M$ slightly larger than 100\,TeV)
we get $F_X\sim 0.3\,$.  Hence, we see that for typical values of the lifetime $\tau_X$,
we can easily suppress the fraction of the $X$ particles that decay within the CRH to $F_X\lsim 0.5\,$.
The long distance decays
of $X$, compared to the Galactic disk scale, ensures that the $\ee$ is dispersed over a larger volume,
which helps diminish the surface brightness of the disk $\ee$ annihilation
emission~\cite{Prantzos:2005pz}.  In addition, this provides a natural mechanism
for transporting  the $\ee$ beyond the CRH, where they could be
guided into the bulge by the large scale poloidal magnetic field.
Hence, the success of case (B) in explaining the
511 keV signal depends on whether enough positrons can be transported into the bulge.
Whether or not this condition is satisfied in our Galaxy is a
question beyond the scope of our paper and poses a challenge to a scenario based on \sncc.

The above discussion outlines how hidden dynamics of the type
discussed in Ref.~\cite{hv1} can in principle produce the gross features
of the observed 511 keV signal.  However, some questions, such as that of the consistency
of this type of scenario with standard cosmology, can only be answered within more specific
models.  As a simple example, we will adopt the One-Light-Flavor (1LF)
model discussed in \Ref{hv1} and mentioned above in section~\ref{micro}.
The key aspects of the phenomenology we are interested in can be captured by considering
only $\eta_h$ and $\omega_h$ and not $\sigma_h$, as we do for simplicity in the following.

\section{Cosmology}\label{cosmology}

In the early universe, once the cosmic temperature $T$ falls below the
dynamical scale $\Lambda_h \sim 1$\,MeV, various hidden hadronic states appear.  The fast interactions
amongst the hidden sector hadrons allow them to annihilate into the lightest state $\eta_h$, as the
universe gets cooler~\cite{hv1}.  If the lightest state has a sufficiently short lifetime, $\tau \ll 1$\,s, then
the hidden sector decays into the SM degrees of freedom in time for BBN.
However, this is generally not the case in our type of scenarios.
Also, a post-BBN hidden hadronic gas could be
unacceptable, as such an ensemble reshifts like matter and would come to dominate the universe
well-ahead of the standard epoch near $T \sim 3$\,eV.  Let us examine the status of the 1LF model
regarding these questions.
From the results in \Ref{hv1}, the lifetimes of $\eta_h$ and $\omega_h$ (assuming that the same
type of expressions are valid for $m \sim 1$\,MeV) have the following dependencies
\beq
\tau_{\eta_h} \propto M^4/\left(f_{\eta_h}^2m_{\eta_h}^5\right) \quad {\rm and} \quad
\tau_{\omega_h} \propto M^4/m_{\omega_h}^5\, ,
\label{tauetaomega}
\eeq
respectively; as mentioned before we will take $f_{\eta_h}\sim m_{\eta_h}$.

\subsection{Case (A)}

With $M\approx 300$\,GeV, $n_c^h = 2$,
and $m_{{\eta_h}, {\omega_h}}\sim 1.5, 2$~MeV, respectively,
we find $\tau _{\eta_h} \sim 10^{21}$~s and
$\tau _{\omega_h} \sim 3$~s.  Here, $\omega_h$
has the kind of lifetime needed for $X$ in case (A).  However,
$\eta_h$ is not sufficiently long-lived to be a safe DM candidate~\cite{Picciotto:2004rp} and
if produced in equilibrium, could come to dominate over radiation
during the BBN, which is not an acceptable outcome.  To examine these questions,
let us estimate the decoupling temperature
of the hidden sector.

Above $T\sim m_X$, we can treat $f$ quarks as free, and their
production in the plasma is governed by their interactions with
$\ee$.  The rate for this interaction is given by $\Gamma_{\rm ef}
\sim n_e \,\sigma_{\rm ef}\, v$. Here, $n_e \sim T^3$, the cross
section $\sigma_{\rm ef}$ is given by \eq{sigma}, but without the
factor of $\alpha/\pi$, the relative velocity $v\sim 1$, and $E_{\rm
cm}\sim 2 T$. The process $\ee \to f {\bar f}$ decouples when
$\Gamma_{\rm ef} < H$, with the Hubble rate $H\simeq 1.7\, g_*^{1/2}
\ T^2/M_{\rm Pl}$; $g_*$ is the number of relativistic degrees of
freedom and $M_{\rm Pl} \simeq 1.2\times 10^{19}$\,GeV is the Planck
mass.  The decoupling temperature, for $g_* \simeq10.8$ ($T\gsim
m_X$) is then given by $T_d \approx 4$\,MeV.  Here, $T_d$ is in
principle high enough for successful BBN.  A similar calculation
also suggests that the $X$ particle, identified in our 1LF example
as $\omega_h$, would be out of equilibrium with the SM plasma for
$T\lsim m_X$.  A simple assumption would be that the reheat
temperature was in the 1-4\,MeV range, such that the $f$-quarks did
not equilibrate with the SM\footnote{A low temperature mechanism for
baryogenesis \cite{Affleck:1984fy,Dimopoulos:1987rk} could provide
the baryon asymmetry of the universe.}. However, arranging for such
a low reheat temperature would require a rather non-standard cosmic
evolution (for bounds on the scale of low temperature inflation
see {\it e.g.}~\cite{Hannestad:2004px} and references therein). We
will next argue that the strong interactions amongst the hidden
sector states could efficiently suppress the relic density of
$\eta_h$ through the relatively fast decays of $\omega_h$, in
equilibrium.

The $\eta_h-\omega_h$ system stays in equilibrium through fast
hadronic interactions governed by a cross section $\sigma v \sim
1/m_{\omega_h}^2$. As the $\omega_h$ population decays, its number
density decreases as
\beq
n(T) \sim T^3 e^{-\Gamma/H},
\label{n}
\eeq
where $\Gamma
\approx 2.1\times 10^{-25}$\,GeV is the width of $\omega_h$, for
$m_{\omega_h}\sim 2$~MeV.  The fast thermal interactions decouple
once $n \sigma v \sim H$, which yields $T_d\sim 0.1$~MeV, and hence
\beq
n(T_d)/T_d^3\sim e^{-\Gamma/H(T_d)}\sim 10^{-20}.
\label{n(Td)}
\eeq
This is a very rough estimate, but shows that the strong hidden dynamics can
efficiently suppress the $\eta_h$ number density to negligible
levels.  We note that a more detailed analysis may be called for to
determine whether the proximity of the BBN era and the onset of
hadronization and decays in the hidden sector does not cause large
deviations from the standard Big Bang picture.  It is important
to note, however, that the $X$ particle density just below the MeV
temperature, when BBN starts, drops very rapidly with
falling temperature, since $n(T)\propto e^{-\Gamma/H}$ and
$\Gamma/H \propto m_X^5/T^2$.  For instance, for
$m_X=3\,$MeV (still in the right range to produce
enough $X$ particles from \snia) and $T=1\,$MeV  we find that $n$ is already a few
percents of its original density which, in this case, is a
small perturbation to the cosmological energy density.

\subsection{Case (B)}

For $M\approx 100$\,TeV and
$m_{{\eta_h}, {\omega_h}}\sim 1.5, 2$\,MeV, we get
$\tau _{\eta_h} \sim 10^{31}$\,s and $\tau _{\omega_h} \sim 10^{11}$\,s.
Thus, given our preceding discussion, we see that $\omega_h$ can easily be a
good candidate for particle $X$, responsible for the \sncc-generated 511-keV signal.  Note also that
$\tau_ {\eta_h}$ is consistent with 511-keV flux bounds~\cite{Picciotto:2004rp},
even if $\eta_h$ is as abundant as DM.  However,
in this case $\eta_h$ will be very long-lived and could upset the
standard picture of cosmology if its relic density is too large.  In case (B), the
hidden sector decouples from
the SM in the early universe at a temperature $T_d \approx 13$\,GeV, where
we have assumed $g_*\simeq86$.
Again, the simplest assumption is that inflation resulted in a
low re-heat temperature of order $T_d$
(within the SM sector) that did not  lead to any
significant production of long-lived hidden hadrons.
Nonetheless, given the longevity of $\eta_h$, it is interesting to see whether there is
a reasonable scenario in which this particle is the dominant DM of the universe.  We
will examine this possibility next.

Let us assume that the primordial SM and hidden sectors evolve in a decoupled
fashion, as discussed above, such that they do not come into thermal equilibrium.
In this case, once the hidden sector's temperature $T_h$ falls below $T_* \sim 1$\,MeV, its
constituents hadronize.  All the heavier hidden hadrons
will quickly annihilate down to $\eta_h$ which will then redshift as matter.  Let us denote the temperature of the
SM sector at this point by $T_i$. In order for matter-radiation equality to take place near its
standard temperature of $T_f \sim 3$\,eV, we then demand
\beq
T_*^4 (T_f/T_i)^3 \sim T_f^4\,,
\label{MREQ}
\eeq
which yields
\beq
10\,{\rm MeV}\lsim T_i \lsim 10^2{\rm\,MeV}
\ \;{\rm for}\; \
 0.3\,{\rm MeV}\lsim T_*\lsim 1{\rm\,MeV}.
\label{TiandT*}
\eeq
We note that the hidden dynamical scale corresponding to $T_*$ can be
somewhat lower than the hadron masses (as in QCD).  Also, over the above range
of $T_i$, the SM and hidden sectors stay decoupled, as $T_d\approx 13$\,GeV.
Hence, in this setup $\eta_h$ can be a
realistic DM candidate, as long as the conditions  (\ref{TiandT*}) are met.  To
complete this discussion, we then suggest a scenario in which these conditions can be realized.
Imagine that the universe went through a period of high scale inflation with a reheat
temperature $T_r \gg T_d$, giving rise to a thermalized plasma
of SM and hidden species.  However, a late and milder inflation can cool this plasma
down to $T\sim 1$\,MeV.  As long as the late inflaton only decays into
SM degrees of freedom and leads to a reheated SM plasma at $T = T_i$, consistent
with (\ref{TiandT*}), $\eta_h$ will survive as a viable DM particle.

\section{Discussion and Summary}\label{summary}

We examined the conditions under which the 511 keV signal could
be related to supernova production of an MeV-scale composite state
$X$ of hidden dynamics in simple models. Our proposed setups
naturally avoid the escape fraction problem of conventional \snia
scenarios.  However, the models we considered face some challenges,
from cosmology or in relation to the spatial distribution of the
signal.  Nonetheless, these proposals offer interesting examples of
how hidden valley dynamics could affect astrophysical observations,
such as that of the 511 keV signal. Our analysis is done at a
semi-quantitative level and a much more detailed study is required
in order to make definite numerical statements.

Nevertheless,
our study demonstrates, in a concrete manner, that hidden valley
dynamics could play an important role in cosmology and astrophysics,
while also having implications for discoveries at accelerators. We
considered how \snia [case (A)] and \sncc [case (B)] explosions can
generate new composite states with relevance to the 511 keV data.

{\it Case (A):} The \snia explosions tend to be
concentrated mostly near the central part of the Galaxy, where the signal
originates.  Conventional explanations based on \snia suffer from the small
escape fraction of $e^+$ generated in radioactive decays of the ejecta.  However,
in our scenario, the $X$ particles have a large enough lifetime
to release the $\ee$ decay products away from the relatively slow moving
explosion front.  Boltzmann suppressed production of $X$ ensures a
non-relativistic emission process and leads to a soft $e^+$ flux.
A minimal setup requires a state $\Phi$ mediating
between the hidden and SM sectors, with a mass $m_\Phi \sim 300$~GeV, weak
coupling to $\ee$ and strong coupling to hidden $f$-quarks.
In this case, $\Phi$ may be directly discovered at a high luminosity
$\ee$ TeV-scale collider.  In addition, in this case, it is interesting that
both type Ia and core collapse supernova dynamics are modified which
may lead to other future signals in precision measurements. If the above effect
has some $z$ dependence it might lead to systematic effects for the next generation
supernova observatories.
Economical hidden valley models can capture the
generic requirements to produce the 511 keV signal and
lead to acceptable cosmic scenarios, though
a more detailed analysis may be required to examine the latter.

{\it Case (B):} An alternative possibility is offered by \sncc.
These explosions are characterized by temperatures far above $m_X$
and can easily provide the required $X$ flux.  The $X$ production is
through hidden sector quark jet fragmentation.  This results in a
soft spectrum for the final state $\ee$, as required.  The requisite
hidden sector production is governed by a large suppression scale
$m_\Phi \sim 100$~TeV, well beyond foreseeable collider reaches,
which leads to long-lived states.  In this case, simple hidden
valley models may potentially include a DM candidate.  Given that
most \sncc explosions take place in the Galactic disk outside the
bulge, the $\ee$ produced in this case must be transported away from
the source.  Achieving the requisite transfer of $e^+$ away from the
disk over such length scales may be a challenge within a
conventional setup.  However, in our scenario, the implied decay
length of $X$ is typically several kpc, allowing for an efficient
deposition of $\ee$ far from the disk.  This leads to a reduced disk
surface brightness by diffusing the $\ee$ over a larger volume and
may help transport of $e^+$ into the bulge, via large scale magnetic
fields of the Galaxy.  These effects could help explain the $B/D$
ratio inferred from observations.  However, the relevance of case
(B) to the 511 keV signal depends on whether the aforementioned
transport mechanism is operative in the Galaxy.


\acknowledgments

We thank M. Strassler for many helpful early discussions, collaboration on topics that led to this project,
and comments on a draft version of the present paper.
We also thank K. Blum, A. Gal-Yam, U. Karshon, C. Lunardini and T. McElmurry  for conversations and comments.
The work of H.D. is supported by the United States Department of Energy
under Grant Contract DE-AC02-98CH10886.
G.P. is supported by the Israel Science Foundation (grant \#1087/09), EU-FP7 Marie Curie, IRG fellowship and the Peter \&
Patricia Gruber Award.


\end{document}